\documentclass[twocolumn,showpacs,preprintnumbers,amsmath,amssymb,groupedaddress,pre]{revtex4}
\usepackage{graphicx}% Include figure files
\usepackage{dcolumn}% Align table columns on decimal point
\usepackage{bm}% bold math
\usepackage{epsfig}

\def\e{\begin{equation}}
\def\f{\end{equation}}
\def\=#1{\overline{\overline #1}}

\def\-#1{{\bf #1}}
\def\.{\cdot}
\def\l#1{\label{eq:#1}}
\def\r#1{(\ref{eq:#1})}
\def\vec#1{{\bf #1}}

\begin{document}

\title{Nonlocal Homogenization Model for a Periodic Array of $\epsilon$-Negative Rods}

\author{M\'ario G. Silveirinha}
\affiliation{Departamento de Eng. Electrot\'{e}cnica da Universidade
de Coimbra, Instituto de Telecomunica\c{c}\~{o}es, P\'{o}lo II, 3030
Coimbra, Portugal}

\date{\today}

\begin{abstract}
We propose an effective permittivity model to homogenize an array of
long thin $\epsilon$-negative rods arranged in a periodic lattice.
It is proved that the effect of spatial dispersion in this
electromagnetic crystal cannot be neglected, and that the medium
supports dispersionless modes that guide the energy along the rod
axes. It is suggested that this effect may be used to achieve
sub-wavelength imaging at the infrared and optical domains. The
reflection problem is studied in detail for the case in which the
rods are parallel to the interfaces. Full wave numerical simulations
demonstrate the validity and accuracy of the new model.

\end{abstract}

 \maketitle

\section{Introduction}

In recent years there has been a lot of interest in the propagation
of electromagnetic waves in artificial materials, and particularly
in materials with negative index of refraction
\cite{Veselago,Shelby}. The research has been mainly driven by the
possibility of using these materials to miniaturize several devices
and waveguides, and develop "perfect lenses" able of focusing
electromagnetic radiation with sub-wavelength resolution
\cite{Pendry_PerfectLens}. Recently, a different approach to achieve
sub-wavelength resolution was proposed in \cite{canal} and
demonstrated experimentally in \cite{SWIWM}. In \cite{SWIWM} the
idea is to use an artificial material formed by an array of
perfectly conducting wires (wire medium) to guide the waves from the
input plane to the output plane, and then reconstruct the image
"pixel by pixel", exploring a Fabry-Perot resonance that in the wire
medium occurs simultaneously for all the spatial harmonics. The
resolution of this transmission device is only limited by the
lattice constant, i.e. by the spacing between the wires. The problem
with the configuration studied in \cite{SWIWM} is that it is limited
to the microwave regime because in the optical domain perfect
electric conducting materials are not available. Nevertheless, we
will suggest in this paper that it may still be possible to use the
same concept to achieve sub-wavelength resolution at the infrared
and optical domains, provided $\epsilon$-negative (ENG) rods are
used to guide the electromagnetic radiation instead of metallic
wires. At the infrared and optical frequencies all metallic
materials have permittivity with negative real part. It is known
that their dielectric constant is well represented by the Drude
model $\epsilon = 1 - \omega_p^2/\omega^2$, where $\omega_p$ is the
plasma frequency (for simplicity the lossless model was considered).
Thus, we envision that either silver or gold rods may used to
fabricate transmission devices that are able to propagate the
subwavelength information of an image.

\begin{figure}[h]
\centering \epsfig{file=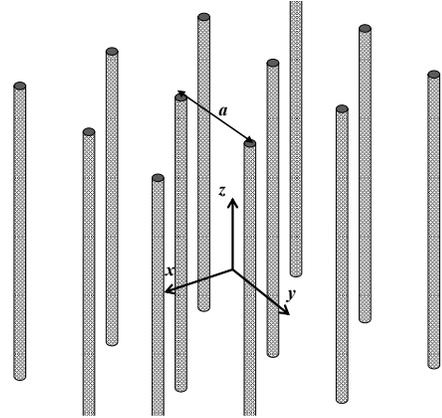, width=6cm} \caption{Medium
formed by long thin ENG rods arranged in a square lattice.}
\label{rods}
\end{figure}

With this motivation, we will investigate in this paper the
propagation of electromagnetic waves in an artificial medium formed
by thin ENG rods (see Fig. \ref{rods}), and we will show that the
structure can be homogenized and described by an effective
permittivity tensor provided spatial dispersion is taken into
account. We will show that our homogenization model predicts that
the rod medium may support a mode that propagates along the axes of
the rods with the same phase velocity, independently of the
transverse wave vector. As proved in \cite{canal}, this is a key
property to operate the material in the canalization regime and
achieve sub-wavelength resolution.

Previous works related with the homogenization of ENG rods are
discussed next. In \cite{Poulton} the oblique propagation of
electromagnetic waves through an array of aligned fibres was
examined, and the associated problem of numerical instability
discussed. In \cite{Pokrovsky_3} a method was proposed to compute
analytically the band structure of wire mesh crystals. In
\cite{Pokrovsky_1} a model was proposed to homogenize a medium with
metallic rods, but only the on-plane case was considered. In
\cite{Pokrovsky_2} an homogenization model was proposed to
characterize a structure similar to the one studied in this paper.
It was demonstrated that the artificial material was characterized
by spatial dispersion, and that the band structure of the photonic
crystal had several branches. However, the results of
\cite{Pokrovsky_2} are restricted to the case in which the
permittivity of the rods follows a Drude-type plasma model, and
besides that the derived formulas for the effective permittivity are
very cumbersome, lead to non-analytical dispersion characteristics,
and more importantly hide the physics of the problem. In this paper,
we will derive a simpler and more intuitive model, describe new
phenomena, and prove that the new approach can characterize
accurately the electrodynamics of the artificial medium.

The homogenization of the wire medium is also closely related with
the subject under study. In fact, the case of perfectly conducting
wires can be regarded as the limit situation in which the
permittivity of the ENG rods is $- \infty $. In \cite{WMPRB} an
homogenization model was derived for the wire medium, and it was
proved that this artificial material suffers from strong spatial
dispersion even for very long wavelengths. Later, in
\cite{Silv_MTT_3DWires, Constantin_WM}, these results were
generalized for 2D- and 3D lattices of connected and unconnected
wires. In \cite{Silv_AP_WM, Pavel_JEA} the reflection problem in a
wire medium slab was investigated, and in \cite{MarioABC} it was
proved that in general an additional boundary condition is necessary
to determine the scattering parameters.

The paper is organized as follows. Firstly, we will characterize the
electric polarizability of a dielectric rod. Then, we will derive
the homogenization rudiments necessary to calculate the effective
permittivity of the structure. In section IV, the electromagnetic
modes supported by the rod medium are characterized. In section V,
we study the reflection problem in a finite slab, assuming that the
rods are parallel to the interfaces. The derived results are
compared with full wave numerical simulations. Finally, in section
VI the conclusions are presented.

In this work we assume that the fields are monochromatic with time
variation $e^{+ j \omega t}$.

\section{Polarizability of a dielectric rod}

Let us consider a ENG rod with radius $R$ and (relative)
permittivity $\epsilon=\epsilon(\omega)$. The rod is oriented along
the $z$-direction. In this section, we calculate the component
$\alpha_{zz}$ of the electric polarizability tensor (actually, since
the rod is infinite along $z$, we are going to calculate the
polarizability per unit of lenght). To this end, we consider that a
plane wave with magnetic field along the $y$-direction illuminates
the dielectric rod. The wave vector of the incident wave is $\vec
k=(k_x,0,k_z)$, where ${\bf{k}}{\bf{.k}} = \beta ^2$ and $\beta
=\omega/c$ is the free-space wave number. The incident electric
field along the $z$-direction is:
\begin{eqnarray}
 E_z^{inc}&&=E_0 e^{ - j\,k_x x} e^{ - j\,k_z z}  \\
  &&=
E_0 e^{ - j\,k_z z} \sum\limits_{n = -\infty}^\infty  { \left( { -
j} \right)^{ \left| n \right|} J_{\left| n \right|} \left( {k_{\rho
,0} r} \right) e^{j n \varphi }  } \nonumber
\l{incidentEz}\end{eqnarray} where $J_n$ is the $J$-Bessel Function
of first kind and order $n$, $k_{\rho ,0} = \sqrt {\beta ^2  - k_z^2
}$, and $(r,\varphi)$ form a system of cylindrical coordinates
attached to the rod axis.

The field components along $z$ can be expanded into cylindrical
harmonics. For example, \e E_z  = \left\{
\begin{array}{l}
 \sum\limits_{n = -\infty}^\infty  {a_n J_{\left| n \right|} \left( {k_{\rho ,m} r} \right) e^{j n \varphi } e^{ - j{\kern 1pt} k_z z} } {\rm{,   }}\,\,r < R  \\
 E_z^{inc}  + \sum\limits_{n = -\infty}^\infty  {b_n H_{\left| n \right|}^{\left( 2 \right)} \left( {k_{\rho ,0} r} \right) e^{j n \varphi } e^{ - j{\kern 1pt} k_z z} } {\rm{,   }}r > R  \\
 \end{array} \right.
\l{Eztotal}\f where $a_n$ and $b_n$ are the unknown coefficients of
the expansion, $H_n^{\left( 2 \right)}=J_n  - jY_n$ is the Hankel
function of second kind and order $n$, and $k_{\rho ,m}  =  - j\sqrt
{k_z^2  - \beta ^2 \varepsilon }$. The field $H_z$ has a similar
expansion.

The transverse fields ${\bf{E}}_{||}$ and ${\bf{H}}_{||}$
(projections into the $xoy$ plane) can be written in terms of $E_z$
and $H_z$:\begin{eqnarray}{\bf{E}}_{||}  = \frac{1}{{k_{\rho ,i}^2
}}\left( { - jk_z \nabla _{||} E_z  + j\beta \,{\bf{\hat u}}_z
\times \nabla _{||} \eta _0 H_z } \right) \l{transv1} \\ \eta _0
{\bf{H}}_{||} = \frac{1}{{k_{\rho ,i}^2 }}\left( { - j\beta
\,\varepsilon _{i} {\bf{\hat u}}_z  \times \nabla _{||} E_z - jk_z
\,\nabla _{||} \eta _0 H_z } \right) \l{transv2}\end{eqnarray} In
the above, $\eta _0$ is the free-space impedance, $\nabla _{||} =
\left( {\frac{\partial }{{\partial x}},\frac{\partial }{{\partial
y}},0} \right)$, $\varepsilon _{i} = \varepsilon$ and $k_{\rho ,i} =
k_{\rho ,m}$ inside the rod, and $\varepsilon _{i} = 1$ and $k_{\rho
,i}  = k_{\rho ,0}$ outside.

The unknown coefficients can be calculated by imposing the
continuity of the tangential electromagnetic fields (i.e. $E_z$,
$E_{\varphi}$, $H_z$, $H_{\varphi}$) at the interface $r=R$.

The ($z$ component of the) electric dipole moment (per unit of
length) is given by,
\begin{eqnarray}
 \frac{{p_z }}{{\varepsilon _0 }} && = \left( {\varepsilon  - 1} \right)\int\limits_{r \le R} {E_z \,ds} \nonumber\\
  && = \left( {\varepsilon  - 1} \right)2\pi R\,a_0 \frac{{J_1 \left( {k_{\rho ,m} R} \right)}}{{k_{\rho ,m} }}e^{ - jk_z z}
\l{pz} \end{eqnarray} and so the polarizability per unit of length
is,\begin{eqnarray}
 \alpha _{zz}  && = \frac{{p_z }}{{\varepsilon _0 \left. {E^{inc} } \right|_{x = y = 0} }}
 \nonumber\\
 && = \left( {\varepsilon  - 1} \right)2\pi R\,\frac{{a_0 }}{{E_0 }}\frac{{J_1 \left( {k_{\rho ,m} R} \right)}}{{k_{\rho ,m} }}
 \l{alfazz} \end{eqnarray}
This result shows that $a_0$ is the unique coefficient required to
calculate the polarizability. Since Maxwell equations are separable
in cylindrical coordinates, to calculate $a_0$ it is sufficient to
impose the boundary conditions to the terms associated with the
cylindrical harmonic $n=0$. It can be easily verified that the $n=0$
term of the magnetic field $H_z$ vanishes (this happens because
$H_z^{inc}=0$ and $\frac{\partial }{{\partial \varphi }} = 0$ for
the $n=0$ harmonic). Thus, only the tangential components $E_z$ and
$H_{\varphi}$ have non-trivial $n=0$ coefficients. Using
\r{incidentEz}-\r{Eztotal} in \r{transv1}-\r{transv2}, and imposing
the continuity of the tangential fields, we find that, \e a_0 J_0
\left( {k_{\rho ,m} R} \right) = b_0 H_0^{\left( 2 \right)} \left(
{k_{\rho ,0} R} \right) + E_0 J_0 \left( {k_{\rho ,0} R} \right)
\l{contEz}\f \e a_0 \frac{\varepsilon }{{k_{\rho ,m} }}J'_0 \left(
{k_{\rho ,m} R} \right) = b_0 \frac{1}{{k_{\rho ,0} }}{H'_0}^{\left(
2 \right)} \left( {k_{\rho ,0} R} \right) + \frac{1}{{k_{\rho ,0} }}
E_0 J'_0 \left( {k_{\rho ,0} R} \right) \f where the prime "$'$"
denotes the derivative of a function. Solving for $a_0$ we readily
obtain:
\begin{eqnarray} a_0^{ - 1}  = && j\frac{{\pi k_{\rho ,0} R}}{2 E_0} \times
 \left( {J_0 \left( {k_{\rho ,m} R} \right){H'_0}^{\left( 2 \right)} \left( {k_{\rho ,0} R} \right)}\right. \nonumber\\ && - \left.{ \frac{{\varepsilon k_{\rho ,0} }}{{k_{\rho ,m} }}J'_0 \left( {k_{\rho ,m} R} \right)H_0^{\left( 2 \right)} \left( {k_{\rho ,0} R} \right)}
 \right) \end{eqnarray}

Substituting this result in \r{alfazz}, we obtain the desired
electric polarizability:
\begin{eqnarray}
 \alpha _{zz}^{ - 1}  = && j\frac{{k_{\rho ,0} }}{4}\frac{1}{{\varepsilon  - 1}} \times
 \left( { - k_{\rho ,m} \frac{{J_0 \left( {k_{\rho ,m} R} \right)}}{{J_1 \left( {k_{\rho ,m} R} \right)}}H_1^{\left( 2 \right)} \left( {k_{\rho ,0} R} \right)}\right. \nonumber \\ && + \left.{ \varepsilon k_{\rho ,0} H_0^{\left( 2 \right)} \left( {k_{\rho ,0} R} \right)} \right)
\l{alfzz_1} \end{eqnarray}

In this work, we consider that the radius of the rods is always much
smaller than the wavelength, or equivalently, $R k_{\rho ,0}<<1$. In
these circumstances, \r{alfzz_1} simplifies to,
\begin{eqnarray}
 \alpha _{zz}^{ - 1}\approx&&
 \frac{1}{{\left( {\varepsilon  - 1} \right)\pi R^2 }} \left[ {1 + j\frac{\pi }{4}\left( {\varepsilon  - 1} \right)\left( {k_{\rho ,0} R} \right)^2 } \right. \nonumber \\
 && \left. { + \frac{1}{2}\left( {\varepsilon  - 1} \right)\left( {C + \log \left( {\frac{{k_{\rho ,0} R}}{2}} \right)} \right)\left( {k_{\rho ,0} R} \right)^2 }
 \right]
\nonumber \l{alfzz_2} \\ \end{eqnarray} where $C$ is the Euler
constant. It is important to note that because the rods are
infinitely long, the polarizability depends not only on the
frequency of the incoming wave, but also on the wave vector
component $k_z$. If $\epsilon$ approaches $-\infty$ the above result
reduces to the case of perfectly conducting rods studied in
\cite{Pavel_JEA}.

Proceeding similarly, we can obtain the well-known formula for the
electric polarizability (per unit of length) in the transverse plane
($xoy$ plane): \e \alpha _{xx}  = \alpha _{yy} \approx
\frac{{\varepsilon - 1}}{{\varepsilon  + 1}}2\pi R^2 \l{alft} \f
Provided the permittivity of the rods is not too close to the
resonance $\epsilon = -1$ and the rods are very thin, the
polarizability can be neglected in the transverse plane.

\section{Effective Permittivity Model}

In the following, we derive an effective permittivity model for the
medium formed by a periodic array of ENG rods. As depicted in Fig.
\ref{rods}, the rods are arranged in a square lattice and the
spacing between the rods (lattice constant) is $a$. As is
well-known, each electromagnetic (Floquet) mode in a periodic medium
can be associated with a wave vector ${\bf{k}}=(k_x, k_y, k_z)$. For
convenience, we define ${\bf{k}_{||}}=(k_x, k_y, 0)$.

To compute the effective permittivity we use the mixing formula : \e
\overline{\overline {\varepsilon }} = \overline{\overline {\bf{I}}}
+ \frac{1}{{{\rm{A}}_{{\rm{cell}}} }}\left( {\overline{\overline
{\alpha }}_e ^{  -1}  - \overline{\overline {{\bf{C}} }
}_{{\mathop{\rm int}}} } \right)^{ - 1} \l{effperm} \f where
${\rm{A}}_{{\rm{cell}}} = a^2$, $\overline{\overline {\bf{I}}}$ is
the identity dyadic, the superscript "-1" represents the inverse
dyadic, and $\overline{\overline {{\bf{C}} } }_{{\mathop{\rm int}}}$
is the interaction dyadic calculated in Appendix B. The mixing
formula is derived in Appendix A, and is valid under the condition
that the dimensions of the cross-section are much smaller than the
lattice constant, and that $|{\bf{k_{||}}}|a<<2\pi$. As discussed in
Appendix A, even though \r{effperm} reminds Clausius-Mossotti
formula, things are not so plain because the lattice has some
intrinsic dispersion. To keep the readability of the paper the
details have been moved to Appendix A.

Next we substitute \r{alfzz_2} and \r{alft} in \r{effperm}. Using
\r{CintStatic}, we find that the effective permittivity in the
transverse plane is \e \varepsilon _{xx}  = \varepsilon _{yy}  = 1 +
\frac{2}{{\frac{1}{{f_V }}\frac{{\varepsilon  + 1}}{{\varepsilon  -
1}} - 1}} \f where $f_V = {{\pi R^2 } \mathord{\left/
 {\vphantom {{\pi R^2 } {{\rm{A}}_{{\rm{cell}}} }}} \right.
 \kern-\nulldelimiterspace} {{\rm{A}}_{{\rm{cell}}} }}
$ is the volume fraction of the rods. Provided the permittivity of
the rods does not satisfy $\epsilon \approx -1$ and the rods are
very thin, we can assume that $\epsilon_{xx} = \epsilon_{yy} \approx
1$. For simplicity, we shall assume this situation in the rest of
the paper. On the other hand, from \r{effperm} it is clear that, \e
\varepsilon _{zz}  = 1 + \frac{1}{{\rm{A}_{{\rm{cell}}}
}}\frac{1}{{\alpha _{zz}^{ - 1}  - C_{{\mathop{\rm int}} ,zz} }}
\l{epsZZ_aux}\f where ${C_{{\mathop{\rm int}} ,zz} }$ is given by
\r{CintFinal}. So, after further simplifications, we obtain that: \e
\varepsilon _{zz} = 1 + \frac{1}{{\frac{1}{{\left( {\varepsilon  -
1} \right)f_V }} - \frac{{\beta ^2  - k_z^2 }}{{\beta _p^2 }}}}
\l{epsfinal}\f where $\beta_p$ is the plasma wave number defined
consistently with the results of \cite{Pavel_JEA} for perfectly
conducting wires:
\begin{eqnarray}
 \left( {\beta _p a} \right)^2 && = \frac{{2\pi }}{{\ln \left( {\frac{a}{{2\pi R}}} \right) + \frac{\pi }{6} + \sum\limits_{n = 1}^\infty  {\frac{1}{{\left| n \right|}}\frac{2}{{e^{2\pi \left| n \right|}  - 1}}} }} \nonumber \\
  && \approx \frac{{2\pi }}{{\ln \left( {\frac{a}{{2\pi R}}} \right) + 0.5275}}
 \l{betp} \end{eqnarray}
Note that in general $\epsilon$ is a function of frequency. Formula
\r{epsfinal} gives the effective permittivity of an array of diluted
ENG rods. The first important observation is that the medium is
spatially dispersive. Indeed the permittivity depends not only on
the frequency $\beta = \omega/c$, but also on the component of the
wave vector parallel to the rods. This means that the medium is
nonlocal, i.e. in the spatial domain the electric displacement
vector and the electric field are related through a spatial
convolution rather than by a multiplication. Secondly, we note that
if the permittivity of the rods approaches $- \infty $, \r{epsfinal}
reduces to the formula derived in \cite{WMPRB} for the wire medium,
consistently with the observation made in the introduction of this
paper. Finally, if we put $\beta=0$ and assume on-plane propagation,
i.e. $k_z=0$, the effective permittivity simplifies to $\varepsilon
_{zz} = 1 + \left( {\varepsilon  - 1} \right)f_V$, which is the
exact formula in the static the limit for non-dispersive dielectrics
\cite{Sihvola}. Thus, very interestingly, formula \r{epsfinal} is in
a certain sense the average of these two limit situations. Note that
even though in this work our main interest is the analysis of ENG
rods, the proposed model is also valid for dielectrics with positive
real part of the permittivity. In the next sections, we characterize
the electromagnetic modes supported by the rod medium and validate
the model with numerical simulations.

\section{Characterization of the electromagnetic modes}

It is evident that the waves in the homogenized medium can be
decomposed into transverse electric (TE) modes (electric field is
normal to the axes of the ENG rods), and transverse magnetic (TM)
modes (magnetic field is normal to the axes of the ENG rods). As
explained in the previous section, we shall assume in this paper
that $\epsilon_{xx}=\epsilon_{yy}\approx 1$ (i.e. that the rods are
very thin, and that the permittivity of rods is not close to
$\epsilon \approx -1$). Within this approximation, the TE-modes do
not interact with the rods, and thus their dispersion characteristic
is, \e \beta ^2  = k^2 \f and the associated average electric field
is (the propagation factor $e^{ - j{\bf{k}}{\bf{.r}}}$ is implicit)
\e {\bf{E}}_{{\rm{av}}}^{TE} \propto \frac{{{\bf{k}}_{||} \times
{\bf{\hat u}}_z }}{{\left| {{\bf{k}}_{||}  \times {\bf{\hat u}}_z }
\right|}} \f On the other hand, the TM-modes satisfy the
characteristic equation: \e k_{||}^2 = \varepsilon _{zz} \left(
{\beta ^2 - k_z^2 } \right) \l{dispTM} \f The above equation cannot
be solved explicitly as a function of $\beta$, because the
permittivity of the rods is itself a function of $\beta$. The
corresponding average electric field is (for $k_z \ne 0$): \e
{\bf{E}}_{{\rm{av}}}^{TM}  \propto \left( {\frac{{{\bf{k}}_{||}
}}{\beta } + \frac{{\beta ^2  - k^2 }}{{\beta ^2 \varepsilon _{zz} -
k^2 }}\frac{{k_z }}{\beta }{\bf{\hat u}}_z } \right) \l{EavTM}\f The
associated magnetic field can be calculated using \r{Max1}.

\begin{figure}[h]
\centering \epsfig{file=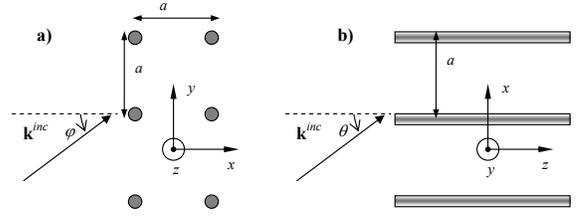, width=8cm} \caption{Reflection of a
plane wave by a semi-infinite rod medium. a)The axes of the rods are
parallel to the interface. b) The axes of the rods are normal to the
interface.} \label{Fig_refl}
\end{figure}

To understand better the nature of the TM-modes, next we study a
reflection problem. Let us consider a semi-infinite rod medium
illuminated from the air side with a plane wave. We analyze two
different geometries, as depicted in Fig. \ref{Fig_refl}.

Firstly, let us suppose that the axes of the rods are parallel to
the interface $x=0$ (Fig. \ref{Fig_refl}$a$). The incident wave
vector is ${\bf{k}}^{inc}=(-j\gamma_0,k_y,k_z)$ with $ \gamma _0 =
\sqrt {k_y^2 + k_z^2  - \beta ^2 } $. It is well-known that the
component of the incident wave vector parallel to the interface,
$(0,k_y,k_z)$, is preserved. In this case only one TM-mode is
excited in the artificial medium (besides the TE-mode). Indeed, from
\r{dispTM} the component $k_x$ of the wave vector inside the rod
medium is given by: \e k_{x,rod}^2 = - k_{y}^2 + \varepsilon _{zz}
\left( {\beta ^2 - k_z^2 } \right) \l{dispTMkx} \f Notice that the
right-hand side of the above equation only depends on the geometry
and parameters of the medium, on the wave number $\beta$ of the
incident wave, and on the components of the incident wave vector
that are preserved $(0,k_y,k_z)$.

Next suppose that the rods are normal to the interface $z=0$ (Fig.
\ref{Fig_refl}$b$). The incident wave vector is now
${\bf{k}}^{inc}=(k_x,k_y,-j\gamma_0)$ with $ \gamma _0 = \sqrt
{k_x^2 + k_y^2  - \beta ^2 } $. The interesting thing is that for
this configuration two TM-modes can be excited inside the rod
medium. Indeed, since ${\bf{k}_{||}}=(k_x,k_y,0)$ to find the
excited electromagnetic modes one needs to solve \r{dispTM} for
$k_z$. Straightforward calculations, using \r{epsfinal}, show that:
\begin{eqnarray} k_z^2  = && \beta ^2  - \frac{1}{2}\left( {\beta _p^2  + k_{||}^2  -
\beta _c^2 }\right.\nonumber \\ && \left.{ \pm \sqrt {\left( {\beta
_p^2 + k_{||}^2 - \beta _c^2} \right)^2 + 4 \beta _c^2 k_{||}^2 } }
\right) \l{dispTMkz}\end{eqnarray} where we defined the parameter
$\beta _c=\beta _c(\omega)$ as, \e \beta _c^2  =  - \frac{{\beta
_p^2 }}{{\left( {\varepsilon \left( \omega  \right) - 1} \right)f_V
}} \l{betc}\f Note that provided the permittivity of the rods is
less than the permittivity of the host medium, $\beta_c$ is a
positive real number (with the same unities as $\beta$; for
simplicity the rods are assumed lossless, otherwise $\beta_c$
becomes a complex number). Also, $\beta_c$ is in general a function
of frequency since $\epsilon$ also is. From \r{dispTMkz} it is seen
that there are two different solutions for $k_z$, and hence two
TM-modes, besides the TE-mode, can propagate inside the artificial
medium. This phenomenon is a manifestation of spatial dispersion,
and is also characteristic of the wire medium \cite{WMPRB}. We also
note that the average electric field for both TM-modes is calculated
using \r{EavTM}.

\begin{figure}[h]
\centering \epsfig{file=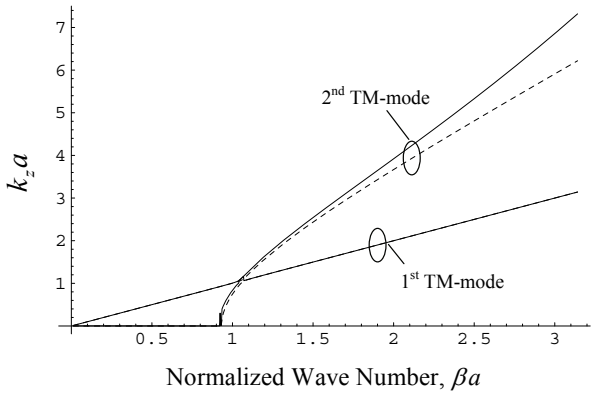, width=8.2cm} \caption{Plot of $k_z$
as a function of normalized frequency $\beta a$, for $R=0.05 a$ and
$k_{||}=0$. The permittivity $\epsilon$ follows a Drude type model
(see the text). The solid line represents the "exact" result,
whereas the dashed line represents the results calculated with the
proposed model.} \label{Fig_disp1}
\end{figure}

To illustrate the discussion, we plot in Fig. \ref{Fig_disp1}, $k_z$
as a function of normalized frequency $\beta a$ for the parameters
$R=0.05 a$, $k_{||}=0$, assuming that the permittivity follows the
Drude model $\epsilon=1- \beta_m^2/\beta^2$ with (normalized) plasma
wave number $\beta_m a = 12.0$. The dashed line curve represents the
results calculated using \r{dispTMkz}. The solid line curve
corresponds to the data calculated by substituting \r{alfzz_1} in
\r{epsZZ_aux} (with no approximations) and calculating
${C_{{\mathop{\rm int}} ,zz} }$ using \r{Cintzz_def} with
${\bf{r}=\bf{r'}}=0$, but without making any assumptions with
respect to ${\bf{k}_{||}}a$ being negligible. Hence, the
permittivity becomes a function of not only $\beta$ and $k_z$, but
also of the other components of the wave vector. The permittivity
function obtained in this way is substituted in \r{dispTM}, and the
corresponding equation is solved numerically. A similar procedure
was used in \cite{Pokrovsky_2, Pavel_JEA}, and so further details
are omitted here. These results can be regarded as "exact" within
the thin rod approximation. As seen in Fig. \ref{Fig_disp1}, the
agreement for the first TM-mode is always excellent. On the other
hand, the second TM-mode is not so accurately predicted for
relatively small wavelengths. The reason is twofold. The first
reason is that for larger frequencies the long wavelength limit
approximation is not so accurate. The second reason will be
discussed later. Fig. \ref{Fig_disp1} shows that for $k_{||}=0$ one
of the electromagnetic modes propagates with the speed of light. It
is also important to refer that for very low frequencies one of the
TM-modes is cut-off (complex imaginary propagation constant). This
is because the ENG rods behave as perfect conductors in the static
limit (when the permittivity follows the Drude model).

To give further insight about the TM-modes, let us study different
limit situations. First, suppose that at some frequency
$\beta_c(\omega)<<\beta_p$, i.e. the permittivity of the rods is
very large in absolute value. In this case, \r{dispTMkz} reveals
that one of the modes has the dispersion characteristic $k_z^2  =
\beta ^2$, and that the other mode has the dispersion $k_z^2  =
\beta ^2 - \beta _p^2  - k_{||}^2$. The former mode can be readily
identified with the well-known transverse electromagnetic (TEM)
dispersionless mode of the wire medium (perfectly conducting wires),
while the latter is the TM-mode of the wire medium.

Consider now the case $\bf{k}_{||}\approx 0$, i.e. paraxial
incidence. Using a Taylor expansion we obtain: \e k_z^2 \approx
\beta ^2  + \frac{{k_{||}^2 }}{2}\left( { - 1 \pm \frac{{\beta _c^2
+ \beta _p^2 }}{{\beta _c^2  - \beta _p^2 }}} \right) + \left\{
\begin{array}{l}
 \beta _c^2  - \beta _p^2  \\
 0 \\
 \end{array} \right. \l{paraxial}
\f The above formula shows that if either $\beta_c(\omega)<<\beta_p$
or $\beta_c(\omega)>>\beta_p$, one of the modes becomes
dispersionless with respect to $\bf{k}_{||}$ (i.e. the coefficient
associated with $k_{||}^2$ vanishes). The former case was already
discussed. As to the latter case, the pertinent mode has dispersion
$k_z^2 \approx \beta ^2  + \beta _c^2$. But this implies that
$k_z>\beta$ and thus this mode is a surface wave guided along the
rods. In \cite{ENG_WG_Tak} it was proved that a ENG rod is able to
support tightly bounded surface modes that propagate electromagnetic
energy with subwavelength beam radius. For metallic materials the
surface modes are surface plasmon polaritons. It was shown that the
energy becomes more confined to the vicinity of the dielectric
waveguide when the effective index of refraction $n_{eff} =
k_z/\beta$ increases. This important result justifies why in the rod
medium one of the TM-modes becomes independent of $\bf{k}_{||}$. In
fact, when $\beta_c(\omega)>>\beta_p$ each guided mode is confined
to a small vicinity of the respective ENG rod, there is no
interaction or coupling between the rods, and consequently one of
the TM-modes becomes dispersionless. As referred before, when
$\beta_c(\omega)<<\beta_p$ there is also a quasi-TEM dispersionless
mode. However this mode is qualitatively very different from the
mode that arises when $\beta_c(\omega)>>\beta_p$. Indeed, while in
the latter case the energy is propagated tightly bounded to the ENG
rods, in the TEM mode case the field energy is distributed more or
less uniformly by the whole volume. As discussed in the
introduction, the dispersionless modes may be used to canalize the
electromagnetic radiation through the rod medium and achieve
sub-wavelength imaging. A detailed analysis of this topic is out of
the scope of the present paper, and will be reported elsewhere. We
also refer that the other TM-mode, still assuming that
$\beta_c(\omega)>>\beta_p$, has to a first approximation the
dispersion characteristic $k_z^2 \approx \beta ^2 - k_{||}^2$, i.e.
approximately the same dispersion as the TE-mode.

\begin{figure}[h]
\centering \epsfig{file=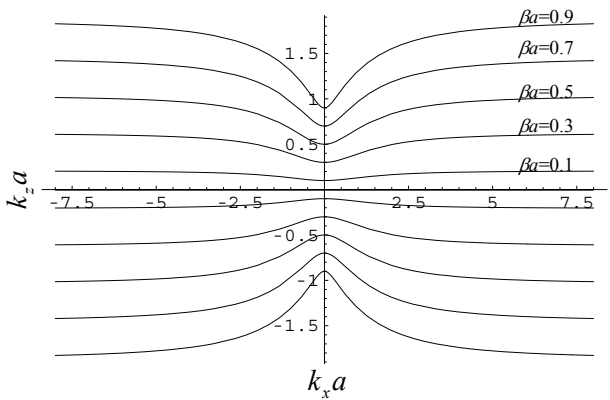, width=8.2cm} \caption{Plot of $k_z$
as a function of $k_x a$ for $R=0.05 a$ and $k_y=0$. Contours $\beta
a = 0.1, 0.3, 0.5, 0.7, 0.9$ for the TM-modes. The permittivity
$\epsilon$ follows a Drude type model (see the text).}
\label{Fig_disp2}
\end{figure}

\begin{figure}[h]
\centering \epsfig{file=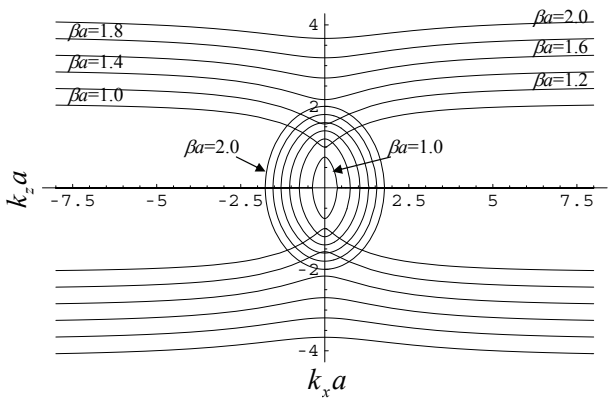, width=8.2cm} \caption{Plot of $k_z$
as a function of $k_x a$ for $R=0.05 a$ and $k_y=0$. Contours $\beta
a = 1.0, 1.2, 1.4, 1.6, 1.8, 2.0$ for the TM-modes. The permittivity
$\epsilon$ follows a Drude type model (see the text).}
\label{Fig_disp3}
\end{figure}

In Figs. \ref{Fig_disp2} and \ref{Fig_disp3}, $k_z$ is plotted as a
function of $k_x$ for $R=0.05 a$ and several different values of the
normalized frequency $\beta a$ (the permittivity of the rods follows
the same Drude model as before). For convenience we show the results
in two different figures. For very long wavelengths
$\beta_c(\omega)<<\beta_p$, and consequently one of the modes is
cut-off. The dispersion of the other mode (shown in Fig.
\ref{Fig_disp2}) becomes increasingly flat as the frequency (and
consequently $\beta_c$) decreases. Note that from \r{betc} and
assuming a Drude type model, $\beta_c$ increases monotonically with
the frequency.

At some point, as the frequency increases, the TM-mode that was
cut-off starts to propagate. In this case, (Fig. \ref{Fig_disp3}),
for each fixed frequency there are two different contours, i.e. two
propagating TM-modes. As the frequency increases even more,
$\beta_c$ becomes comparable or larger than $\beta_p$, and the band
structure of one of the TM-modes becomes practically flat,
consistently with the fact that the energy propagated by this mode
is tightly confined to the vicinity of the ENG rods.

In Figs. \ref{Fig_disp2} and \ref{Fig_disp3} it can also be seen
that around $k_{||} \approx 0$ the wave normal contours of one of
the modes are to some extent hyperbolic. In fact, using \r{paraxial}
it can be easily checked that one of the modes has dispersion
characteristic such that the signs of the coefficients associated
with $k_{||}^2$ and $k_z^2$ are symmetric. This property is more
important for frequencies such that $\beta_c(\omega) \approx
\beta_p$. Note that if the medium was anisotropic, with no spatial
dispersion, and negative permittivity along the $z$-direction the
contours would also be hyperbolic. It is well-known that hyperbolic
contours may originate negative refraction at an interface.

It is also important to refer that in the limit $\varepsilon \to 1$,
the ratio $\beta_c(\omega) / \beta_p$ becomes infinitely large (see
 \r{betc}) and consequently $k_z / \beta$ is also infinitely
large. This means that in these circumstances the localized TM-mode
either cannot be excited by an incoming wave, or if it is excited it
is killed by losses. Thus, only the other TM-mode, with dispersion
$k_z^2 \approx \beta ^2 - k_{||}^2$, will propagate. This is
consistent with the fact that as $\varepsilon \to 1$ the medium
shall have the same properties as free-space.

To conclude this section, we will discuss the scope of application
of the permittivity model \r{epsfinal}. We remember that the results
were derived for thin rods, $R<<a$, and under the assumption that
$|k_{\rho ,0}| a << \pi$ and $\beta a << \pi$. In general the modes
that propagate in the long wavelength limit satisfy the previous
conditions without problems. However, there is one exception at the
problem at hand. In fact, when $\beta_c(\omega)>>\beta_p$ the radial
constant $k_{\rho ,0}$ becomes complex imaginary for the TM-mode
associated with the surface mode (surface plasmon). As discussed in
\cite{ENG_WG_Tak}, when the beam radius is subwavelength the
effective index of refraction $n_{eff} = k_z/\beta$ becomes very
large, and in that case the condition $|k_{\rho ,0}| a << \pi$ may
not be observed. This situation affects the accuracy of our model
when $\beta_c(\omega)>>\beta_p$. Indeed, the error in the TM-mode
associated with the surface mode becomes non-negligible in this
situation.  The other TM-mode is still accurately predicted. This
result justifies deterioration of the agreement in Fig.
\ref{Fig_disp1}, as the frequency (and consequently, for the Drude
Model, also $\beta_c$) increases.

Fortunately, it is easy to solve this problem. In fact, when
$\beta_c(\omega)>>\beta_p$ the dispersion characteristic of the
pertinent TM-mode is essentially the same as the dispersion
characteristic of the guided mode supported by a single ENG rod.
This dispersion characteristic is determined in \cite{ENG_WG_Tak},
and is equivalent to the condition $\alpha _{zz}^{ - 1}=0$, where
$\alpha _{zz}^{ - 1}$ is given by \r{alfzz_1}. Thus, to summarize
our findings, the TM-modes can be accurately calculated using
\r{dispTMkz}, except when $\beta_c(\omega)>>\beta_p$ which yields
less accurate results for the mode with higher $k_z$. In this case
the corresponding TM-mode is dispersionless, and follows the same
characteristic as the guided mode supported by a single dielectric
rod.

\section{The Reflection Problem with rods parallel to the interfaces}

\begin{figure}[h]
\centering \epsfig{file=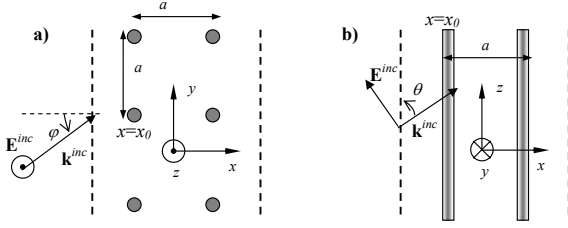, width=8.2cm} \caption{A plane
wave illuminates a slab of the rod medium ($N_L=2$). The rods are
parallel to the plane $x=0$. a) $k_z=0$. b) $k_y=0$.}
\label{Fig_reflslab}
\end{figure}

To further validate the proposed permittivity model, we will study
the reflection of electromagnetic waves by a rod medium slab with
finite thickness. We will suppose that the rods are parallel to the
interface (see Fig. \ref{Fig_reflslab}). The slab consists of $N_L$
layers of rods. The structure is periodic in $y$ and $z$, and the
dielectric rods stand in free-space. The interfaces $x=x'_L$ and
$x=x'_R = x'_L + N_L a$ are represented by the dashed lines (since
the rods stand in air the definition of the interfaces is a bit
ambiguous; this will be discussed below with more detail). The rods
in the leftmost layer are in the plane $x=x_0$.

As discussed in the previous section, when the rods are parallel to
the interface, equation \r{dispTMkx} has only one solution for
$k_x^2$. For simplicity, we will restrict our attention to the case
in which either $k_z = 0$ (Fig. \ref{Fig_reflslab}a) or $k_y = 0$
(Fig. \ref{Fig_reflslab}b). For these particular geometries, an
incident plane wave polarized as depicted in Fig. \ref{Fig_reflslab}
can only excite the TM-mode inside the rod medium. As is well-known
\cite{Collin}, in the general case where $k_y$ and $k_z$ are
simultaneously different from zero, both the TM- and TE-modes are
excited (the medium is birefringent). Apart from the more heavy
notation, the general case poses no additional difficulties.

Using \r{Max1}, \r{EavTM}, and \r{dispTMkx} and matching the
tangential components of the electric and magnetic fields at the
interfaces, we find that the reflection coefficient referred to the
plane $x=x'_L$ is given by:
\begin{eqnarray} \rho  = \frac{{\tanh \left( {\gamma _m d} \right)\left(
{\gamma _0^2 - \gamma _m^2 } \right)}}{{2\gamma _0 \gamma _m  +
\tanh \left( {\gamma _m d} \right)\left( {\gamma _0^2  + \gamma _m^2
} \right)}} \l{roa}
\\
\rho  = \frac{{ \tanh \left( {\gamma _m d} \right)\left(
{\varepsilon ^2 \gamma _0^2  - \gamma _m^2 } \right)}}{{2\varepsilon
\gamma _0 \gamma _m + \tanh \left( {\gamma _m d} \right)\left(
{\varepsilon ^2 \gamma _0^2  + \gamma _m^2 } \right)}} \l{rob}
\end{eqnarray}
where $d=N_L a$ is the thickness of the slab, $\gamma _m  =
j\,k_{x,rod}$, $\gamma _0 = \sqrt {k_y^2 + k_z^2  - \beta {}^2}$,
and \r{roa} corresponds to Fig. \ref{Fig_reflslab}a) with $k_z=0$,
and \r{rob} corresponds to Fig. \ref{Fig_reflslab}b) with $k_y=0$.

\begin{figure}[h]
\centering \epsfig{file=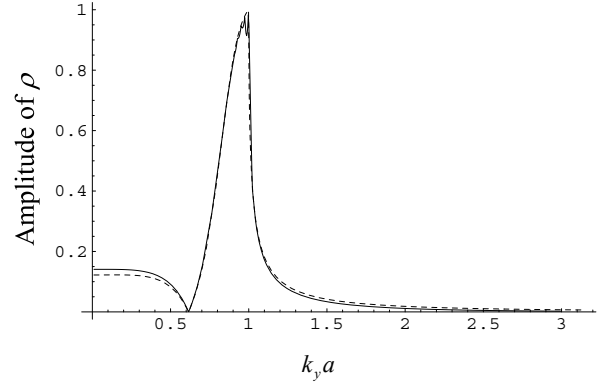, width=8.0cm} \caption{Reflection
coefficient as a function of the wave vector component parallel to
the interface. The slab consists of $N_L = 5$ layers, and the rods
have $R=0.05a$, and $\epsilon = -30$ at $\beta a = 1.0$  The solid
line represents the full wave MoM data, and the dashed line
represents the data obtained using the analytical model.}
\label{Fig_Ex1}
\end{figure}

Next, in order to demonstrate the accuracy of the theoretical
results, the analytical model is tested against full wave data
computed with the periodic moment method (MoM) \cite{Wu}. In the
first example we consider that $R=0.05 a$, and $\epsilon=-30.0$ at
$\beta a= 1.0$ (for simplicity, losses are neglected). A plane wave
polarized as depicted in Fig. \ref{Fig_reflslab}a illuminates 5
layers of rods ($N_L = 5$). The amplitude of the reflection
coefficient is depicted in Fig. \ref{Fig_Ex1} as a function of $k_y
a$. Note that the angle of incidence $\varphi$ is such that  $ \sin
\varphi
  = {{k_y } \mathord{\left/
 {\vphantom {{k_y } \beta }} \right.
 \kern-\nulldelimiterspace} \beta }
$. For $k_y > \beta$ the incident wave is evanescent. The solid line
represents the MoM full wave data. The dashed line represents the
results computed using the proposed permittivity model (formula
\r{roa}). It is seen that that the agreement between the two sets of
data is good. Similar agreement is obtained for the phase of the
reflection coefficient and for the transmission coefficient. The
previous results also demonstrate that the homogenization model is
useful to study not only incident propagating plane waves, but also
part of the evanescent spectrum.

\begin{figure}[h]
\centering \epsfig{file=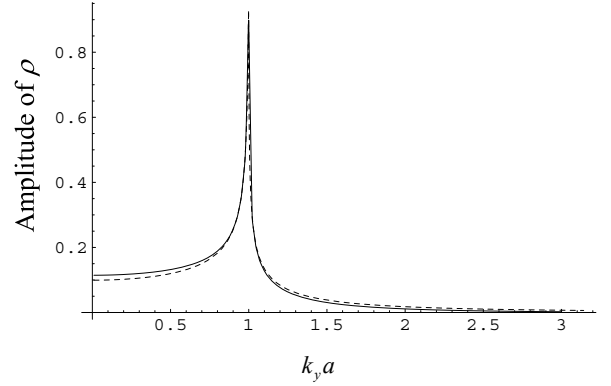, width=8.0cm} \caption{Reflection
coefficient as a function of $k_y a$ for $N_L = 1$ layer, $R=0.05a$,
and $\epsilon = -30$ at $\beta a = 1.0$. The solid and dashed lines
are defined as in Fig. \ref{Fig_Ex1}.} \label{Fig_Ex2}
\end{figure}

As noted before, since the rods stand in free-space the position of
the interfaces and thickness of the slab are a bit ambiguous. Notice
that the thickness of the homogenized slab was taken equal $d=N_L
a$, apparently with good results. Next, to test if this choice still
yields accurate results for very thin slabs, the reflection
coefficient is computed for the same structure, except that now the
slab has only one layer of rods ($N_L = 1$). The reflection
coefficient is depicted in Fig. \ref{Fig_Ex2} and has a peak at $k_y
a =1.0$, which corresponds to the transition between propagating
waves and evanescent waves. As seen, even though the slab is so thin
the agreement is still remarkably good. This is a bit a surprising,
because for such a thin slabs one would expect that the interface
effects and granularity of the artificial medium would prohibit the
homogenization of the structure using the bulk medium average
fields.

\begin{figure}[h]
\centering \epsfig{file=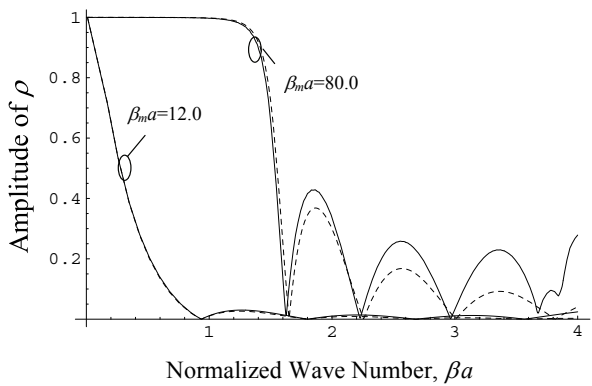, width=8.0cm} \caption{Reflection
coefficient as a function of the free-space wave number $\beta a$
for $R=0.01 a$, and $N_L = 5$ layers. The permittivity of the rods
follows a Drude-type model (see the text). The solid and dashed
lines are defined as in Fig. \ref{Fig_Ex1}. \label{Fig_Ex3}}
\end{figure}

In the next example, we study what happens if the angle of incidence
$\varphi$ is kept constant ($\varphi = 45 [\rm{deg}]$), and the
frequency is varied. Now $R=0.01 a$, $N_L = 5$, and the permittivity
follows the Drude model $\epsilon=1- \beta_m^2/\beta^2$ with plasma
wave number $\beta_m a = 12.0$ or $\beta_m a = 80.0$. The calculated
results are shown in Fig. \ref{Fig_Ex3}. For relatively low
frequencies the two sets of data agree very well, but as the
frequency increases the agreement progressively deteriorates, since
the long wavelength limit approximation is no longer valid. Notice
that for relatively low frequencies the medium blocks the incident
radiation because the rods effectively behave as perfectly
conducting wires.

\begin{figure}[h]
\centering \epsfig{file=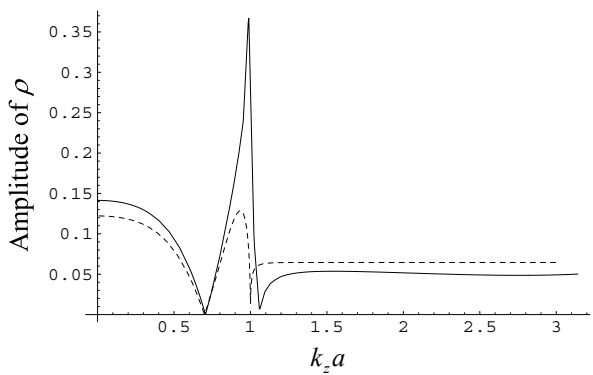, width=8.0cm} \caption{Reflection
coefficient as a function of $k_z a$ for $N_L = 5$ layer, $R=0.05a$,
and $\epsilon = -30$ at $\beta a = 1.0$. The solid and dashed lines
are defined as in Fig. \ref{Fig_Ex1}. \label{Fig_Ex4}}
\end{figure}

So far the wave vector of the incident wave was always perpendicular
to the axes of the ENG rods, and so the effects of spatial
dispersion were hidden. In the last example, we consider the very
different propagation scenario depicted in Fig. \ref{Fig_reflslab}b.
The parameters of the rods are $R=0.05a$, $\epsilon = -30$, and $N_L
= 5$. The reflection coefficient for $\beta a = 1.0$ is shown in
Fig. \ref{Fig_Ex4} as a function of $k_z a$. Note that the angle of
incidence $\theta$ for propagating waves satisfies $ \sin \theta =
{{k_z } \mathord{\left/
 {\vphantom {{k_z } \beta }} \right.
 \kern-\nulldelimiterspace} \beta }
$. Fig. \ref{Fig_Ex4} shows that the agreement between the numerical
and analytical results is still quite satisfactory, except near the
transition between the propagating waves and the evanescent waves ($
k_z a \approx 1.0$).

It is important to refer that the configurations studied in this
section (rods parallel to the interface) are not appropriate to
achieve sub-wavelength imaging. For that application the rods must
be perpendicular to the interface, as shown in Fig. \ref{Fig_refl}b.
As discussed in section IV, for this geometry three modes are
excited in the (homogenized) rod medium, and thus an additional
boundary condition is necessary to solve the scattering problem. The
same situation occurs in the wire medium \cite{MarioABC}. The
analysis of this problem is out of the scope of this work and will
be presented elsewhere.

\section{Conclusion}

We studied the electrodynamics of a periodic array of thin ENG rods.
Using the polarizability of a single rod and integral equation
methods, we derived new nonlocal permittivity model for the
artificial medium that accurately describes the propagation of waves
in the long wavelength limit. We discussed the effects of spatial
dispersion in the context of the reflection problem. It was proved
that when the rods are parallel to interface only two modes (TE and
TM) can be excited in the artificial medium. However, when the axes
of the rods are normal to the interface, two TM-modes, besides the
TE- mode, can be excited in the medium, as a manifestation of
spatial dispersion. It was demonstrated that the wave normal
contours of one of the TM-modes are intrinsically hyperbolic. It was
proved that the rod medium supports dispersionless modes that
propagate along the axes of the rods, and it was speculated that
this important property may allow sub-wavelength imaging of
electromagnetic waves at the infrared and optical domains, using the
idea proposed in \cite{canal}. It was shown that the energy of the
dispersionless modes can be loosely bounded to the ENG rods (in this
case the wave is essentially transverse electromagnetic), or
alternatively tightly confined to a vicinity of the rods. In the
latter case, the modes have approximately the same dispersion as the
guided surface mode supported by an individual rod. The reflection
problem was investigated in detail for the case where rods are
parallel to the interfaces. The developed theory was successively
tested against full wave data calculated with the MoM.

\begin{acknowledgments}
The author thanks Pavel A. Belov for valuable discussions.
\end{acknowledgments}

\appendix
\section{Derivation of the Mixing Formula}

In this Appendix we derive the mixing formula \r{effperm} used to
homogenize the rod medium.

Let us consider a generic electromagnetic Floquet mode
$(\bf{E},\bf{H})$, associated with the wave vector ${\bf{k}} =
(k_x,k_y,k_z)$ and the wave number $\beta =\omega/c$, i.e. the
fields satisfy Maxwell-Equations and $({\bf{E}},{\bf{H}}) \exp
\left( {{j} {\bf{k}}{\bf{.r}}} \right)$ is a periodic function in
the lattice. The average electric field is defined as, \e
{\bf{E}}_{{\rm{av}}} = \frac{1}{{{\rm{V}}_{{\rm{cell}}}
}}\int\limits_\Omega {{\bf{E}}\left( {\bf{r}} \right)e^{ +
j\,{\bf{k}}{\bf{.r}}} } d^3 {\bf{r}} \l{Eavdef} \f and
${\bf{H}}_{{\rm{av}}}$ is defined similarly. In the above,  $\Omega$
represents the unit cell of the periodic medium. From
Maxwell-Equations it can be proved that:
\begin{eqnarray}
 - {\bf{k}} \times {\bf{E}}_{{\rm{av}}}  + \beta \,\eta _0 {\bf{H}}_{{\rm{av}}}  = 0
\l{Max1} \\
\beta \,\left( {{\bf{E}}_{{\rm{av}}}  + \frac{{\bf{P}}}{{\varepsilon
_{\rm{0}} }}} \right) + {\bf{k}} \times \eta _0 {\bf{H}}_{{\rm{av}}}
= 0 \end{eqnarray} where the generalized polarization vector is
given by, \e \frac{{\bf{P}}}{{\varepsilon _0 }} =
\frac{1}{{{\rm{V}}_{{\rm{cell}}} }}\int\limits_\Omega  {\left(
{\varepsilon  - 1} \right){\bf{E}}\,} e^{ + j\,{\bf{k}}{\bf{.r}}}
d^3 {\bf{r}} \f Straightforward calculations show that these
relations imply that: \e {\bf{E}}_{{\rm{av}}}  = \frac{{\beta ^2
\overline{\overline {\bf{I}}}  - {\bf{kk}}}}{{k^2  - \beta ^2
}}{\bf{.}}\frac{{\bf{P}}}{{\varepsilon _0 }} \l{EavPol} \f The above
equations are exact and completely general. Next we will apply these
results to the rod medium under study. To begin with, we note that
since the crystal is invariant to translations along the
$z$-direction the fields depend on $z$ as $\exp ( - j k_z z)$. Using
standard Green function methods \cite{Collin}, it can be proved that
the electric field has the following integral representation, \e
{\bf{E}}\left( {\bf{r}} \right) = \left( {\varepsilon - 1}
\right)\int\limits_{S} {\beta ^2 \overline{\overline {{\bf{G}} }}_p
\left( {\left. {\bf{r}} \right|{\bf{r'}}} \right){\bf{.E}}\left(
{{\bf{r'}}} \right)} \,\,d^2 {\bf{r'}} \l{fieldrep}\f where the
primed and unprimed coordinates represent the source and observation
points, respectively, $S = \left\{ {\left( {x',y',0} \right):x'^2  +
y'^2  \le R^2 } \right\}$ is the cross-section of the dielectric rod
in the unit cell, and the Green function dyadic is defined by, \e
 \overline{\overline {{\bf{G}} }}_p  =
 \left( {\overline{\overline {\bf{I}}}  + \frac{1}{{\beta {}^2}}\left( {\nabla _{||}  - jk_z {\bf{\hat u}}_z } \right)\left( {\nabla _{||}  - jk_z {\bf{\hat u}}_z } \right)} \right)\,\Phi _p
\f In the above, $\Phi _p = \Phi _p\left( {\left. {\bf{r}}
\right|{\bf{r'}}} \right)$ is the dynamic potential created by a
phase-shifted array of line sources, \e \nabla^2 \Phi _p + \beta^2
\Phi _p = - \sum\limits_{\bf{I}} {\delta \left( {{\bf{r_{||}}} -
{\bf{r'_{||}}} - {\bf{r}}_{\bf{I}} } \right)} e^{ -
j{\bf{k}}{\bf{.}}\left( {{\bf{r}}  - {\bf{r'}} } \right)} \l{HEq} \f
where $\delta$ is the Dirac function, ${\bf{I}}=(i_1,i_2)$ is a
double index of integers, ${\bf{r_I}}=a(i_1, i_2, 0)$ is a lattice
point, and $r_{||} = (x,y,0)$. Thus $\Phi _p$ is intrinsically
two-dimensional, depending on $z$ and $z'$ as $e^{ - jk_z \left( {z
- z'} \right)}$. Furthermore, it is obvious that the Green function
only depends on the relative position $\bf{r_{||}} - \bf{r'_{||}}$.
Note that the Green function can be written as a superimposition of
the potentials created by the line sources: \e \Phi _p  =
\sum\limits_{\bf{I}} {\Phi _0 \left( {{\bf{r}}_{||} - {\bf{r'}}_{||}
- {\bf{r}}_{\bf{I}} } \right)} e^{ - j{\bf{k}}{\bf{.r}}_{\bf{I}} }
\l{spatRep}\f where $\Phi _0  = e^{ - jk_z \left( {z - z'} \right)}
{{H_0^{\left( 2 \right)} \left( {k_{\rho ,0} \left| {{\bf{r}}_{||} -
{\bf{r'}}_{||} } \right|} \right)} \mathord{\left/
 {\vphantom {{H_0^{\left( 2 \right)} \left( {\beta \left| {{\bf{r}}_{||}  - {\bf{r'}}_{||} } \right|} \right)} {4j}}} \right.
 \kern-\nulldelimiterspace} {4j}}$ is the potential created by a
 line source placed at $\bf{r'_{||}}$, i.e. the solution of
 \r{HEq} when the summation in the right-hand side is restricted
 to the index ${\bf{I}} = {\bf{0}}$. Physically, \r{fieldrep}
 establishes that the electric field at some point of space is the
 superimposition of the fields radiated by all the dielectric rods
 of the lattice.

The Green potential can also be written as a Fourier series since it
is a (pseudo) periodic function of the wave vector: \e \Phi _p
\left( {\left. {\bf{r}} \right|{\bf{r'}}} \right) =
\frac{1}{{{\rm{A}}_{{\rm{cell}}} }}\sum\limits_{\bf{J}} {\frac{{e^{
- j{\bf{k}}_{\bf{J}} {\bf{.}}\left( {{\bf{r}}  - {\bf{r'}} }
\right)} }}{{{\bf{k}}_{\bf{J}} {\bf{.k}}_{\bf{J}}  - \beta ^2 }}}
\l{specRep}\f where $ {\rm{A}}_{\rm{cell}} = a^2$, ${\bf{J}} = (j_1,
j_2)$ is a double index of integers, $ {\bf{k}}_{\bf{J}} = {\bf{k}}
+ {\kern 1pt} \,{\bf{k}}_{\bf{J}}^0$, and ${\bf{k}}_{\bf{J}}^0 =
{{2\pi } \mathord{\left/
 {\vphantom {{2\pi } a}} \right.
 \kern-\nulldelimiterspace} a}\left( {j_1 ,j_2, 0 } \right)$.

 Now that the necessary theoretical formalism was introduced, we are
 ready to study the homogenization problem in the rod
 medium. To begin with, we note that from \r{spatRep} and \r{specRep} the Green potential
 is singular in the spatial domain, i.e. when $\bf{r_{||}} - \bf{r'_{||}}
 \approx 0$ (source region), as well as in the spectral domain, i.e. when
${\bf{k}}{\bf{.k}} \approx \beta ^2 $ (long wavelength
 limit). Since the integral \r{fieldrep} is defined over the source
 region and we want to study the electromagnetic modes that propagate in the long wavelength limit, it is convenient to single
 out the terms that make the Green function singular and decompose
 it as follows:
 \e
\Phi _p  = \Phi _0  + \frac{1}{{{\rm{A}}_{{\rm{cell}}} }}\frac{{e^{
- j{\bf{k}}{\bf{.}}\left( {{\bf{r}}  - {\bf{r'}} } \right)}
}}{{{\bf{k}}{\bf{.k}} - \beta ^2 }} + \Phi _{reg} \l{phidecomp} \f
where $\Phi _{reg}$, which is defined implicitly by the above
formula, is a regular function both in the spatial domain (source
region $\bf{r_{||}} - \bf{r'_{||}} \approx 0$) and in the spectral
domain (long wavelength limit). Using this decomposition in
\r{fieldrep} we find that:
\begin{eqnarray}
 {\bf{E}}\left( {\bf{r}} \right) = && {\bf{E}}_{{\rm{av}}} e^{ - j{\bf{k}}{\bf{.r}}}  + \left( {\varepsilon - 1} \right)\int\limits_{S} {\beta ^2 \overline{\overline {{\bf{G}} }}_0 \left( {\left. {\bf{r}} \right|{\bf{r'}}} \right){\bf{.E}}} \,\,d^2 {\bf{r'}} \nonumber \\
  && + \left( {\varepsilon - 1} \right)\int\limits_{S} {\beta ^2 \overline{\overline {{\bf{G}} }}_{reg} \left( {\left. {\bf{r}} \right|{\bf{r'}}} \right){\bf{.E}}} \,\,d^2 {\bf{r'}}
 \l{fieldrep2} \end{eqnarray}
where ${\bf{E}}_{{\rm{av}}}$ is the average field in the crystal,
and $\overline{\overline {{\bf{G}} }}_{0}$ and $\overline{\overline
{{\bf{G}} }}_{reg}$ are defined as $\overline{\overline {{\bf{G}}
}}_{p}$, except that the Green potential is replaced by $\Phi _{0}$
and $\Phi _{reg}$, respectively. To obtain the above result we used
\r{EavPol}. Next, we use the fact that the ENG rods are assumed to
be very thin ($R/a<<1$), and that we want to investigate the
electrodynamics of modes that propagate in the long wavelength limit
($|{\bf{k_{||}}}|a<<2\pi$ and $\beta a<<2\pi$). Since the dyadic
$\overline{\overline {{\bf{G}}}}_{reg}$ is regular in both the
spatial and spectral domains, it is legitim to write (putting
$\bf{r_{||}}=\bf{r'_{||}}=0$ and ${\bf{k_{||}}}=(k_x, k_y, 0)=
\bf{0}$), \e \overline{\overline {{\bf{G}} }}_{reg} \left( {\left.
{\bf{r}} \right|{\bf{r'}};{\bf{k}},\beta } \right) \approx
\overline{\overline {{\bf{G}} }}_{reg} \left( {z|z';k_z,\beta }
\right) \f being the formula valid in the source region. For
convenience, we introduce the following interaction dyadic: \e
\overline{\overline {{\bf{C}} } }_{{\mathop{\rm int}}}  = \beta ^2
\overline{\overline {{\bf{G}} }}_{reg} \left( {0;k_z,\beta } \right)
\l{intDyad}\f Then, it is clear from \r{fieldrep2} that,
\begin{eqnarray}
 {\bf{E}}\left( {\bf{r}} \right) \approx &&
\left( {{\bf{E}}_{{\rm{av}}}  + \overline{\overline {{\bf{C}} }
}_{{\mathop{\rm int}}} .\frac{{{\bf{p}} }}{{\varepsilon _0 }}}
\right)e^{ - jk_z z}
\nonumber \\
  && + \left( {\varepsilon - 1} \right)\int\limits_{S} {\beta ^2 \overline{\overline {{\bf{G}} }}_{0} \left( {\left. {\bf{r}} \right|{\bf{r'}}} \right){\bf{.E}}} \,\,d^2 {\bf{r'}}
 \l{field_source} \end{eqnarray}
provided the observation point $\bf{r}$ is near the source region
and $|{\bf{k_{||}}}|a<<2\pi$. In above, we introduced the electric
dipole moment (per unit of length), ${\bf{p}}$, of the dielectric
rod in the unit cell. Now, the key result is that \r{field_source}
is formally equivalent to the (integral) equation obtained when a
single rod is illuminated by a plane wave with electric field with
amplitude ${\bf{E}}^{{\rm{inc}}} = \left( {{\bf{E}}_{{\rm{av}}}  +
\overline{\overline {{\bf{C}} } }_{{\mathop{\rm int}}}
.\frac{{{\bf{p}} }}{{\varepsilon _0 }}} \right)$ and the same wave
vector component $k_z$ along the $z$-direction. In other words, when
the rod in the unit cell stands alone in free-space and is
illuminated with the above defined plane wave, the total electric
field also satisfies (to a first approximation) \r{field_source} in
the source region. But this remarkable result implies that: \e
\frac{{\bf{p}}}{{\varepsilon _0 }} = \overline{\overline
{{\bf{\alpha }} }}_e {\bf{.}}\left( {{\bf{E}}_{{\rm{av}}}  +
\overline{\overline {{\bf{C}} } }_{{\mathop{\rm int}}}
{\bf{.}}\frac{{\bf{p}}}{{\varepsilon _0 }}} \right) \f where
$\overline{\overline {{\bf{\alpha }} }}_e$ is the electric
polarizability tensor for a single rod. The term inside brackets in
the right-hand side can be regarded as the local field that
polarizes a single rod embedded in the dielectric crystal. Within
the thin rod condition and for long wavelengths, the above solution
is exact.

We are now ready to calculate the effective permittivity dyadic.
Since the (macroscopic) electric displacement vector $\bf{D}$ is
given by ${\bf{D}} = \varepsilon _0 {\bf{E}}_{{\rm{av}}}  +
{{\bf{p}} \mathord{\left/
 {\vphantom {{\bf{p}} {{\rm{A}}_{{\rm{cell}}} }}} \right.
 \kern-\nulldelimiterspace} {{\rm{A}}_{{\rm{cell}}} }}
$, and the effective permittivity tensor must guarantee that $
{\bf{D}} = \varepsilon _0 \overline{\overline {\varepsilon }}
{\bf{.E}}_{{\rm{av}}}$, we conclude that the effective permittivity
of the rod medium is given by the mixing formula \r{effperm}. Note
that \r{effperm} is completely general and is valid independently of
the specific geometry of the transverse section of the rod.

At this point it is appropriate to compare \r{effperm} with the
classic homogenization approach. It is striking that \r{effperm}
reminds Clausius-Mossotti formula \cite{Collin, Sihvola}. Indeed, if
we could identify the interaction dyadic $\overline{\overline
{{\bf{C}} } }_{{\mathop{\rm int}}}$ with $ {1 \mathord{\left/
 {\vphantom {1 {2{\rm{A}}_{{\rm{cell}}} }}} \right.
 \kern-\nulldelimiterspace} {2{\rm{A}}_{{\rm{cell}}} }}
$ the formulas would be the same (note also that the rods are
arranged in a square lattice). In Appendix B we calculate the
interaction dyadic in closed-form using the static limit
approximation. Equation \r{CintStatic} shows that the interaction
dyadic is different from $ {1 \mathord{\left/
 {\vphantom {1 {2{\rm{A}}_{{\rm{cell}}} }}} \right.
 \kern-\nulldelimiterspace} {2{\rm{A}}_{{\rm{cell}}} }}
$ only along the $z$-direction. This important result shows that
Clausius-Mossotti formula is not valid for media with cylindrical
inclusions. More specifically it fails to predict the effective
permittivity along the direction in which the crystal is uniform.
This is an indirect manifestation of spatial dispersion.

We also mention that the interaction dyadic defined by \r{intDyad}
is not equivalent to the dynamic interaction constant defined in
other works (see for example \cite{Pavel_PRE_05}). Indeed, in our
definition we extracted the singularities in both the spatial and
spectral domains, while other works usually only extract the
singularity in the spatial domain. It is clear from our previous
analysis that it is the singularity in the spectral domain that
indirectly defines the relation between the local field that
polarizes the rod and the average field.

\section{Calculation of the interaction dyadic}

Here we calculate the interaction dyadic defined by \r{intDyad}. It
can be written as: \e \overline{\overline {{\bf{C}} }
}_{{\mathop{\rm int}}}  = \left( {\beta ^2 \overline{\overline
{\bf{I}}}  + \left( {\nabla _{||}  - jk_z {\bf{\hat u}}_z }
\right)\left( {\nabla _{||} - jk_z {\bf{\hat u}}_z } \right)}
\right) \Phi _{reg} \l{Cint2} \f where the right-hand side of the
expression is evaluated at $\bf{r}=\bf{r'}=0$ and $\bf{k_{||}}=0$.
From \r{HEq} and \r{phidecomp} it is clear that:
\begin{eqnarray}
 && \nabla^2 \Phi _{reg} + \beta^2 \Phi _{reg} = \nonumber \\ && \left(
\frac{1}{{{\rm{A}}_{{\rm{cell}}} }} - \sum\limits_{{\bf{I}} \ne
{\bf{0}}} {\delta \left( {{\bf{r_{||}}} - {\bf{r'_{||}}} -
{\bf{r}}_{\bf{I}} } \right)} \right) e^{ - j{\bf{k}}{\bf{.}}\left(
{{\bf{r}}  - {\bf{r'}} } \right)}
\end{eqnarray}
Putting $\bf{r}=\bf{r'}=0$ and $\bf{k}=0$ in the above equation, and
letting $\beta$ approach zero (static limit), we find that: \e
\left. {\nabla ^2 \Phi _{reg} \left( {{\bf{r}} = {\bf{r'}} =
0;{\bf{k}} = {\bf{0}}} \right)} \right|_{\beta  = 0}  =
\frac{1}{{{\rm{A}}_{{\rm{cell}}} }} \l{nab2phi} \f Moreover, because
of the symmetry of the square lattice it is evident that if
$\bf{r}=\bf{r'}=0$ and $\bf{k}=0$ the following relations hold, $
\frac{{\partial ^2 \Phi _{reg} }}{{\partial x_i
\partial x_j }} = 0 $ if $ i \ne j$, $ \frac{{\partial ^2 \Phi
_{reg} }}{{\partial z^2 }} = 0 $, and $ \frac{{\partial ^2 \Phi
_{reg} }}{{\partial x^2 }} = \frac{{\partial ^2 \Phi _{reg}
}}{{\partial y^2 }} $. So using \r{nab2phi}, we conclude that in the
static limit ($\bf{k}=0$ and $\beta=0$) the interaction dyadic is
given by: \e \overline{\overline {{\bf{C}}} }_{{\mathop{\rm int}}} =
\frac{1}{{2{\rm{A}}_{{\rm{cell}}} }}\left( {\overline{\overline
{\bf{I}}}  - {\bf{\hat u}}_z {\bf{\hat u}}_z } \right)
\l{CintStatic} \f The above result is consistent in the $xoy$ plane
with the (two dimensional version of the) Clausius-Mossotti formula.
However, along the $z$-direction, maybe a bit surprisingly, the
interaction constant vanishes in the static limit. Next, we will
estimate ${C_{{\mathop{\rm int}} ,zz} }$ in the dynamic case. From
\r{Cint2}, we have that: \e C_{{\mathop{\rm int}} ,zz}  = \left(
{\beta ^2  - k_z^2 } \right)\Phi _{reg} \l{Cintzz_def}\f Using
\r{spatRep} and \r{phidecomp}, and putting $\bf{r}=\bf{r'}=0$ and
$\bf{k}_{||}=0$, we obtain that: \e C_{{\mathop{\rm int}} ,zz} =
\left( {\beta ^2  - k_z^2 } \right)\sum\limits_{{\bf{I}} \ne 0}
{\frac{1}{{4j}}H_0^{\left( 2 \right)} \left( {k_{\rho ,0} \left|
{{\bf{r}}_{\bf{I}} } \right|} \right)}  +
\frac{1}{{{\rm{A}}_{{\rm{cell}}} }} \f The series in the right-hand
side was evaluated in \cite{Pavel_JEA}. Using the results of
\cite{Pavel_JEA}, and assuming that $k_{\rho ,0} a << \pi$, we
obtain:
\begin{eqnarray}
 C_{{\mathop{\rm int}} ,zz}  \approx && k_{\rho ,0}^2 \left( {\frac{j}{4} + \frac{1}{{2\pi }}\ln \left( {\frac{{k_{\rho ,0} a}}{{4\pi }}} \right) + \frac{C}{{2\pi }} + } \right. \nonumber \\
&& \left. { + \frac{1}{{12}} + \sum\limits_{n = 1}^\infty
{\frac{1}{{\pi \left| n \right|}}\frac{1}{{e^{2\pi \left| n \right|}
- 1}}} } \right) \l{CintFinal} \end{eqnarray} where $C$ is the Euler
constant.

\bibliography{rods}
\end{document}